# Drop impact on wet granular beds: water-content effects on the cratering

**Wei Zhang**[1]
張 犇

**Hiroaki Katsuragi**[1]
桂木 洋光

**Ken Yamamoto**[*1,2]
山本 憲

[1] Department of Earth and Space Science, Osaka University, 1-1 Machikaneyama, Toyonaka, Osaka 560-0043, Japan
[2] Water Frontier Research Center (WaTUS), Tokyo University of Science, 6-3-1 Niijuku, Katsushika-ku, Tokyo 125-8585, Japan

[*] Corresponding author: yam@ess.sci.osaka-u.ac.jp

**Abstract**
Drop impact events on wet granular bed show rich variety by changing the substrate composition. We observe the drop impact onto dry/wet granular substrates with different grain size (50–400 μm) and water content (0–22 vol %). Although the impactor condition is fixed (impact velocity: 4.0 m s$^{-1}$, water drop radius: 1.8 mm), the experiment reveals that the post-impact behaviors of both impactor and target are strongly affected by the substrate composition. We sort these behaviors into several phases regarding liquid splashing and crater shapes left after the event. As these phases show relevance each other, we measure the mechanical characteristics of the substrates and find that the onset of splashing and particle ejection are explained by a fracture of the substrate. Furthermore, we discuss several timescales of the event to understand more detailed phase separations. Consequently, we find that the splashing phase and the crater shape are determined by a competition of the timescales of impact, penetration, and contact.

## I. INTRODUCTION

The spherical beauty of liquid drops has attracted numerous scientists and engineers. Even in our everyday lives, one can easily find a lot of spherical liquid drops. Examples include dew, mist, and milk crown formation. Due to the gravity, relatively large liquid drops tend to fall and impact on the floor. The target floors are sometimes hard solids, liquid pools at other times, and granular matter such as soil at still other times. The liquid drop impact onto various target materials, therefore, has been extensively studied thus far[1,2]. Particularly, splashing induced by the drop impact has been studied by varying surrounding pressure[3,4], target elasticity[5–8], and target surface structure[9–14], etc. Recently, the splashing modes have been classified into two types: prompt splashing and corona splashing[2,15,16]. The drop impact phenomenon is one of the most active research topics in the field of fluid-related physics and engineering, and has been studied extensively. Nevertheless, the mechanics of the impacting drop deformation and breaking are still under active debate.

    When the target consists of a collection of solid particles, observed phenomena become much more complex because of the deformation and ejection of the granular target. By the drop impact on a granular bed, various crater shapes and rebounding modes have been found[17–22]. For instance, scaling for crater diameter and/or depth has been frequently discussed to characterize the crater shape[17,23,24].

    The drop impact cratering relates to soil erosion by raindrop impact. Therefore, detailed analyses of the crater formation and splashing have been performed recently in the field of agro-physics[25–27]. To discuss the natural soil erosion process, complex effects of grain shape, grain-size variation, water content, etc. have to be carefully considered. This is contrastive to the fundamental physical studies in which ideal situation (*e.g.*, dry, monodispersed, and spherical glass beads) is usually employed. For example, while Zhang *et al.* examined the

effect of slightly wet granular target[24], systematic variation of water content has not been studied. However, very wet situation is crucial to mimic actual soil conditions.

Crater shapes formed on complex wet granular targets relate also to planetary problems. Various types of peculiar crater shapes relating to wet granular impact have been reported[28–31]. However, systematic experiments to reveal the details of geologically observed complex craters have not been carried out so far.

Under certain conditions, the liquid drop impact onto a granular surface results in a liquid marble, a water drop covered by a thin granular layer. Since liquid marble has the potential for various applications, its fundamental studies have been extensively conducted recently[32]. The liquid marble formation must be revealed to properly understand the physical origin of the variation of the crater shapes observed in the drop impact onto a granular target. The systematic investigation of the drop impact onto a granular surface is a pressing issue also from the view-point of both fundamental and applicational understanding of the liquid marbles and peculiar crater shapes.

Based on the above-mentioned background, we focus on the effects of water content and grain size on the drop–granular impact phenomena, in this study. Both the grain size and water content of the target granular bed are systematically varied. Instead, drop size and impact velocity are fixed. Besides, only spherical glass beads are used to concentrate on the effects of grain size and water content. Particularly, onset conditions for the drop splashing and grain ejection are measured based on high-speed imaging. To explain these observed onset conditions, the mechanical properties of the target granular bed are also measured by the indentation test. In addition, the morphology of the resultant final crater shapes is measured and analyzed. Using these data, the onset conditions of drop splashing and ejector release are linked to the effective strength and event timescales of the target granular bed and resultant final crater shapes.

## II. EXPERIMENTAL SETUP AND PROCEDURE

### A. Impactor and target

We release a water drop with a radius $R_0$ of 1.8 mm from a flat-tipped needle locating 80 cm above the target substrate. The drop slowly grows at the edge of the needle owing to a flow supplied by a syringe pump *via* connecting tube at low infusion rate (0.1 mL/min), and leaves the needle tip when the gravitational force exceeds the surface tension force. The drop then freefalls and impacts onto the substrate with the impact velocity $U_0$ of 4.0 m s$^{-1}$. The resultant Weber number $We$ and Reynolds number $Re$ are $We = \rho_w U_0^2 R_0 / \gamma = 395$ and $Re = \rho_w U_0 R_0 / \eta = 7186$, respectively, where $\rho_w$, $\gamma$, and $\eta$ denote water density, water–air interfacial tension, and water viscosity, respectively.

The target granular bed substrate is composed of monodispersed spherical glass beads (density of 2500 kg m$^{-3}$) and water. We prepare various substrates with four different grain diameters ($d_g$ = 50, 100, 200, and 400 μm) and seven different water contents [$w$ = 0 (dry), 0.62, 1.2, 2.5, 4.9, 12, and 22 vol %]. The water content is defined by the total volume of substrate $V_{substrate}$ and that of added water $V_{water}$ as $w = V_{water} / V_{substrate}$. We prepare the wet target by adding water to dry granular at a specified volume ratio and stirring it to achieve uniform water content in the substrate. The target is then loaded in a lab-made container, which has a cylindrical hole (20 mm in diameter and 55 mm in depth). The drop-impact experiment is performed immediately after the target preparation in order to avoid a non-uniform-water-content condition due to the drainage. We repeated experiments of the same condition at least three times to confirm the reproducibility.

### B. Measurements

*High-speed observation*

We observe the impact events from an obliquely upward direction using a high-speed camera (SA-5, Photron) with a macro lens (AF Micro-Nikkor 200 mm f/4D IF-ED, Nikon) with a backlight illumination by an LED light source (LA-HDF158A, Hayashi Watch-works). The event is recorded at 10000 fps with a spatial resolution of 25 μm / pixel.

*3D-profile measurement*

3D profiles of the target substrates before and after the event are measured using a laser profilometer. The laser profilometer is composed of a line-scanning laser sensor head (LJ-V7080, Keyence) connected to a controller (LJ-V7000, Keyence) and a stepper-motor-driven translation stage (PM80B-200X, COMS) controlled by a position controller (CP-310, COMS). The line-scan frequency and the field-sweep velocity are set at 20 Hz and 1 mm s$^{-1}$, respectively. The spatial resolutions in horizontal and vertical directions are 5 μm and 10 μm, respectively. We note that we use relative-height profiles (profiles whose height is obtained by subtracting the height before the event from that after the event) because our substrates have nonnegligible roughness stemming from the components.

*Indentation test*
Changing substrate components would bring differences in mechanical characteristics of the target. Although the dynamic characteristics should represent the impact event, we measure the static characteristics by the indentation test using a precision universal testing machine (AG-X, Shimadzu) to characterize the substrate response. The test is performed by inserting a test rod (diameter $D_{\rm rod}$ of 10 mm) into the substrate at a constant velocity of 0.1 mm min$^{-1}$. Time evolution of the insertion depth and stress are recorded for further analyses.

## III. RESULTS

Single water-drop-impact events were captured by the high-speed camera from obliquely upward angle. All the events have the same impactor conditions ($U_0$ = 4.0 m s$^{-1}$, $R_0$ = 1.8 mm) and various target conditions ($d_{\rm g}$ = 50, 100, 200, and 400 μm, $w$ = 0–22 vol %). We also measured 3D profiles of craters left after the impact. Consequently, we observed a rich variety of the post-impact phenomena as well as the crater shapes (Fig. 1) and divided them into several phases from perspectives of the liquid splashing, the particle ejection, and the crater shape as shown in Fig. 2. We describe the characteristics of these phases in the following subsections.

### A. Liquid splashing

We observed the prompt splash[2] and receding breakup[33], and found a tendency that the prompt splash is dominant for small $d_{\rm g}$ and high $w$. We note that we refer the splashes such that shown in Fig. 1a, e, and f as the prompt splash because the small droplets are released directly from the advancing lamella[2], which, in this study, took off the substrate.

It is also noteworthy that we also observed a small amount of tiny liquid fragments (less than 100 μm in diameter) at very early stage of the event (~0.5 ms after the impact) for $d_{\rm g} \geq 100$ μm (see blue dashed line in Fig. 2a). Besides, the takeoff angle of the fragment splash showed significant variation depending on the degree of penetration of the drop into the substrate, and it was close to 90° for the deepest penetration ($d_{\rm g}$ = 400 μm, $w$ = 22 vol %). We excluded this type of splashing from the splashing phase diagram shown in Fig. 2 for clarity.

The splashing was divided into four phases as follows with respect to the difference in the splash angle, the degree of penetration, and the existence of the prompt/receding splash (see Fig. 2).

- Phase 1 (Fig. 1e): prompt splash on the substrate (in the horizontal plane). This phase was only observed at $w \geq 2.5$ vol % with $d_{\rm g}$ = 50 μm. The spreading front of the liquid is directed parallel to the substrate. Only the prompt splash is possible owing to the contact of the liquid front with the substrate, which is similar to the spread on hydrophilic substrates[34]. The particles stay in the substrate during the event.
- Phase 2 (Fig. 1a, f): prompt splash in upward direction. The spreading front of the liquid is redirected to upward owing to the change in the substrate morphology (cratering) in early stage of the impact and it takes off the substrate. The contactless liquid-front breaks up into small droplets by interface instabilities as it is observed on flat rigid substrates, while minimizing the viscous drag stemming from the liquid–solid contact[13,35]. The upward angle of the splash increases with $d_{\rm g}$ and decreases with $w$. The spreading liquid film, which either does not contain particles or contains a small number of particles, ruptures in air.
- Phase 3 (Fig. 1b, g): weak receding breakup along with deep penetration. The drop deeply penetrates into the substrate and then spread while generating ejecta particles. The spreading interface traps significant number of particles at the same time. It finally breaks up into liquid-marble-like droplets.
- Phase 4 (Fig. 1c, d, h): no splash. The drop deeply penetrates into the substrate and generates ejecta particles, but does not break up into small droplets.

It is also remarkable that Phase 3 and Phase 4 cover almost all region where the ejecta particles were observed (Fig. 2a), which is intuitively understandable from the fact that the degree of the penetration was qualitatively small for Phase 1 and Phase 2.

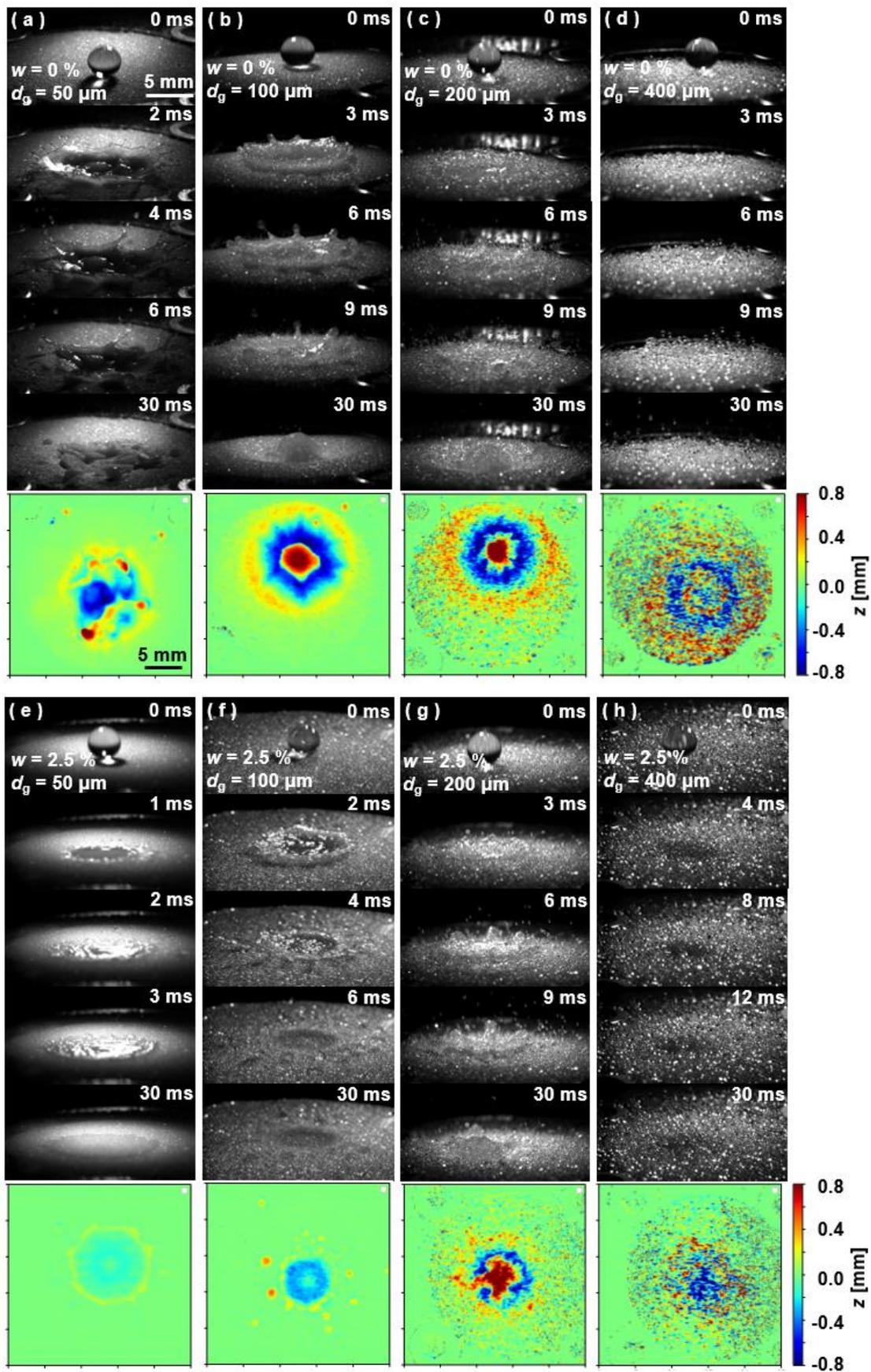

FIG. 1. Image sequences and crater 3D profiles of selected events. (a) $d_g$ = 50 μm, $w$ = 0 vol % (Phase 2). (b) $d_g$ = 100 μm, $w$ = 0 vol % (Phase 3). (c) $d_g$ = 200 μm, $w$ = 0 vol % (Phase 4). (d) $d_g$ = 400 μm, $w$ = 0 vol % (Phase 4). (e) $d_g$ = 50 μm, $w$ = 2.5 vol % (Phase 1). (f) $d_g$ = 100 μm, $w$ = 2.5 vol % (Phase 2). (g) $d_g$ = 200 μm, $w$ = 2.5 vol % (Phase 3). (h) $d_g$ = 400 μm, $w$ = 2.5 vol % (Phase 4). Scale bars indicate 5 mm.

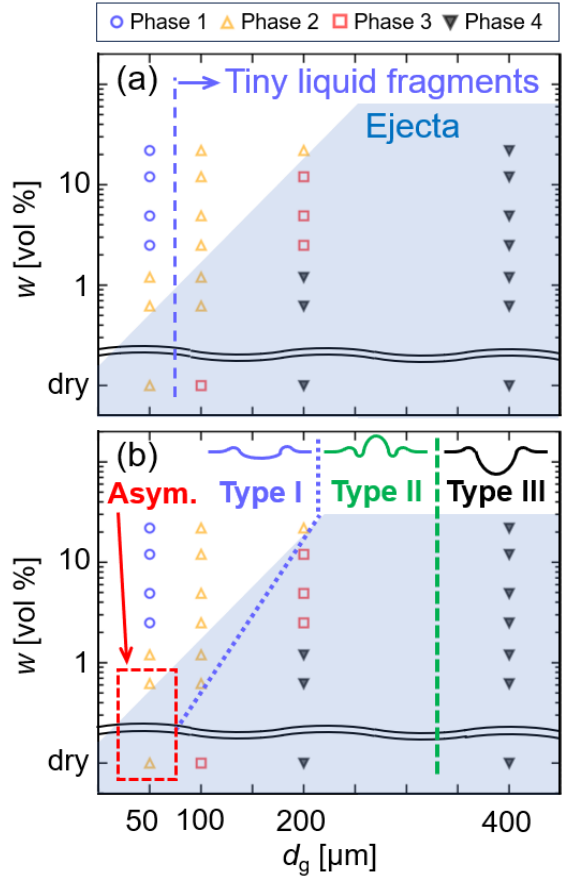

FIG. 2. Phase diagrams of the post-impact event. Liquid splashing is divided into four phases; Phase 1 (blue open circle): prompt splash in horizontal direction, Phase 2 (orange open upper triangle): prompt splash in upward direction, Phase 3 (red open square): deep penetration & weak receding splash, Phase 4 (black filled lower triangle): deep penetration without splash. (a) The diagram also indicates the boundaries of the tiny liquid fragments (< 100 μm) generation (observed for $d_g$ > 50 μm) and the particle ejection (ejecta observed in the blue-grayed area). (b) Boundaries for three crater types (Type I: shallow flat craters, Type II: a dome in center, Type III: deep bowl-like craters) with schematics of the cross-section are depicted as well as the boundary of the asymmetric craters (red-dashed rectangle).

## B. Crater profiles

The ejecta particles were observed (the blue-grayed area in Fig. 2, see also Fig. 1a–d, g, and h) when the impactor (water drop) penetrated into the substrate, and a crater was left behind. The crater shapes were almost axi-symmetric except for the cases of $d_g$ = 50 μm, $w$ = 0 and 0.62 vol % (Fig. 1a), but showed several variations as it is noticeable in both the last sequencies and 3D profiles shown in Fig. 1. To characterize the crater shape, we obtained azimuthally averaged cross-sectional profiles ($r$-$z$ plane, where $r$ and $z$ denote radial and vertical directions, respectively) as shown in Fig. 3. One finds that, even when the substrate is wet, a round-shaped rim is formed at the outskirt of the crater as on the dry granular. However, the profiles reveal that the water content affects the crater shape. We sort them into three types based on the shape inside the crater rim (Fig. 2b): (I) inner-rim is almost flat and slightly sagged from the original level, (II) a raised dome at the center and a deep gutter around the dome are formed, and (III) a bowl-like deep depression is formed. The center dome of Type II craters is composed of particles that were trapped in the spreading drop and gathered by the retraction of the liquid front. In case of the Type III crater formation, the impacting drop deeply penetrates into the substrate and does not widely spread in horizontal direction, which results in the bowl-like deep hole formation at the center. It is interesting that the boundary between Type I and Type II corresponds to that between Phase 2 and Phase 3 of splashing, while the boundary between Type II and Type III locates in between $d_g$ = 200 and 400 μm.

In contrast to the crater morphology, the crater rim radius $R_c$ and the maximum depth $z_c$ (defined in Fig. 4a) do not show remarkable difference for different conditions, but they tend to be slightly large when $w$ is low as shown in Fig. 4b and c. However, a reverse tendency is found in $z_c$ for $d_g$ = 400 μm, presumably because of the different crater shape (Type III).

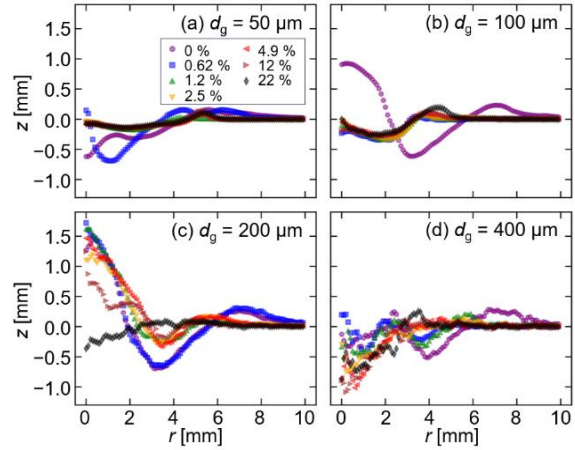

**FIG. 3.** Azimuthally averaged cross-sectional profiles for different water content $w$. (a) $d_g = 50$ μm, (b) $d_g = 100$ μm, (c) $d_g = 200$ μm, and (d) $d_g = 400$ μm.

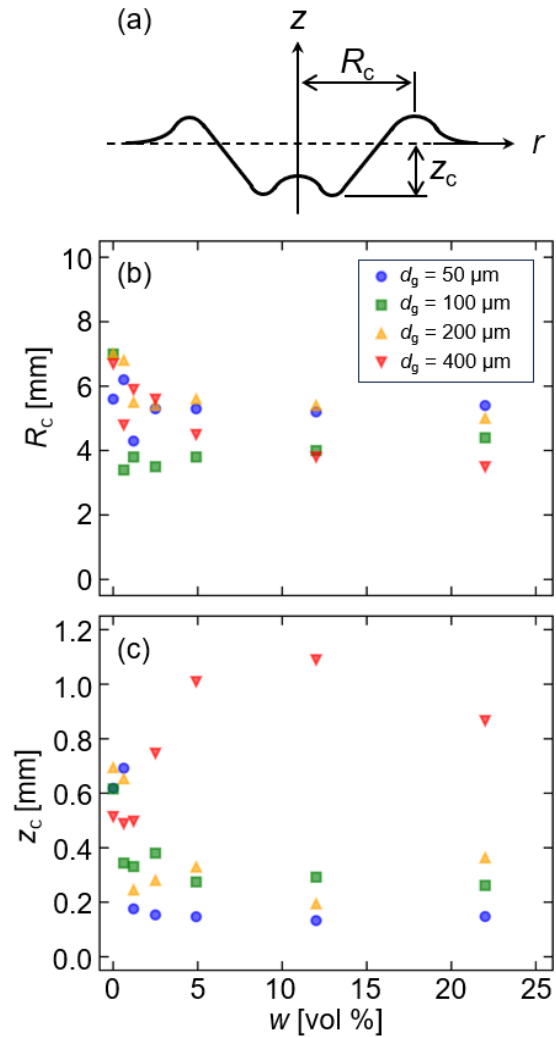

**FIG. 4.** (a) A schematic of the crater cross-section with definitions of the crater rim radius $R_c$ and the maximum depth $z_c$. (b) $R_c$ and (c) $z_c$ as a function of the water content $w$.

## C. Mechanical characteristics of the substrates

We measured the root-mean-square ($RMS$) and $R_a$ roughness of the substrate from the 3D profiles taken before the impact (Table 1). As a result, both values increased with the grain diameter but they did not show clear dependence on the water content $w$. The result is not intuitive because one would expect that the substrates become smoother when they contain liquid. It could be because of the water drainage, which is nonnegligible for $d_g > 100$ μm[36].

We also evaluated the mechanical characteristics of the substrate by the indentation test. Figure 5 shows correlations between the penetration depth of the test rod $S$ and the pressure from the substrate $P$, which was obtained by dividing the measured force by the rod cross-sectional area. There are tendencies that the pressure is high for small $d_g$ and high $w$. However, the tendencies become unclear for $d_g = 200$ and 400 μm.

The relationships shown in Fig. 5 cannot be directly interpreted such like the results obtained with a general elastic solid because the substrates are composed of grains and their relocation should be taken into account in this study. Therefore, we define the effective elasticity $E_{eff}$ and the effective strength $Y_{eff}$ from the measured pressure as follows:

$$E_{eff} = dP/d\varepsilon|_{max} \quad (1)$$

$$Y_{eff} = P(t_b) \quad (2)$$

where $\varepsilon$ denotes the distortion defined as $\varepsilon = S / D_{rod}$ and $t_b$ is the time at which $dP/d\varepsilon = E_{eff} / 2$ after reaching $dP/d\varepsilon|_{max}$. The obtained $E_{eff}$ and $Y_{eff}$ are plotted against the water content $w$ in Fig. 6a and b, respectively. We can confirm that both of them tend to increase with $w$, and the substrate elasticity increases to a level of hard plates ($\sim 10^6$ kPa[7,37]) only by the addition of a small amount of water for $d_g = 50$ and 100 μm. However, for $d_g = 200$ and 400 μm, both $E_{eff}$ and $Y_{eff}$ do not show an increasing trend in $w > 1$ vol %. It can be considered as a result of the water drainage. Moreover, Fig. 6c indicates that $E_{eff}$ and $Y_{eff}$ has a power-law correlation ($E_{eff} \sim Y_{eff}^{0.82}$) in the present system. The same diagram in a normalized form (normalized by the dynamic pressure $P_d = 0.5\,\rho_w U_0^2$) for the abscissa is shown in Fig. 6d.

Table 1. Root-mean-square roughness ($RMS$) and average roughness ($R_a$) measured from 3D profiles. The table also contains the average values over different water content (Avg.) and its standard deviation (SD).

| $d_g$ [μm] | 50 μm | | 100 μm | | 200 μm | | 400 μm | |
|---|---|---|---|---|---|---|---|---|
| $w$ [vol %] | $RMS$ [um] | $R_a$ [um] | $RMS$ [um] | $R_a$ [um] | $RMS$ [um] | $R_a$ [um] | $RMS$ [um] | $R_a$ [um] |
| 0 | 7.2 | 5.7 | 47.6 | 5.0 | 824.1 | 135.8 | 1548.7 | 480.4 |
| 0.62 | 2.8 | 2.1 | 74.8 | 4.8 | 645.7 | 83.6 | 1606.0 | 515.0 |
| 1.2 | 2.6 | 2.0 | 67.0 | 4.9 | 718.9 | 103.5 | 1712.0 | 585.5 |
| 2.5 | 2.3 | 1.8 | 66.8 | 3.4 | 746.4 | 111.5 | 1627.4 | 529.9 |
| 4.9 | 3.7 | 2.8 | 58.0 | 2.8 | 685.3 | 93.9 | 1674.9 | 560.1 |
| 12 | 2.4 | 1.8 | 94.5 | 4.3 | 642.1 | 82.5 | 1621.5 | 524.8 |
| 22 | 2.6 | 2.0 | 82.0 | 4.6 | 702.2 | 98.7 | 1754.8 | 615.5 |
| Avg. | 3.4 | 2.6 | 70.1 | 4.3 | 709.3 | 101.4 | 1649.3 | 544.5 |
| SD | 1.7 | 1.4 | 15.5 | 0.8 | 63.0 | 18.4 | 69.5 | 45.7 |

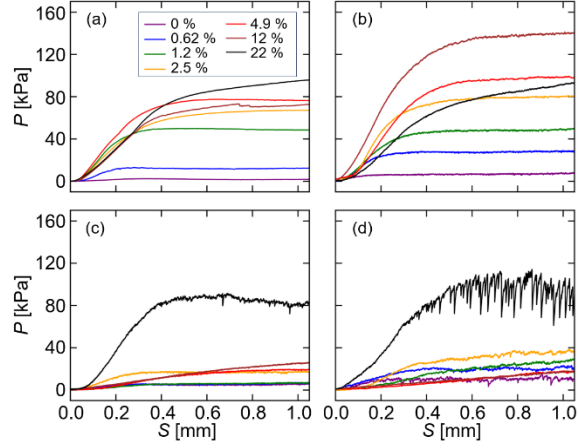

**FIG. 5.** Correlations between the test-rod penetration depth $S$ and pressure $P$ for different water content $w$ obtained by the indentation test. (a) $d_g = 50$ μm, (b) $d_g = 100$ μm, (c) $d_g = 200$ μm, and (d) $d_g = 400$ μm.

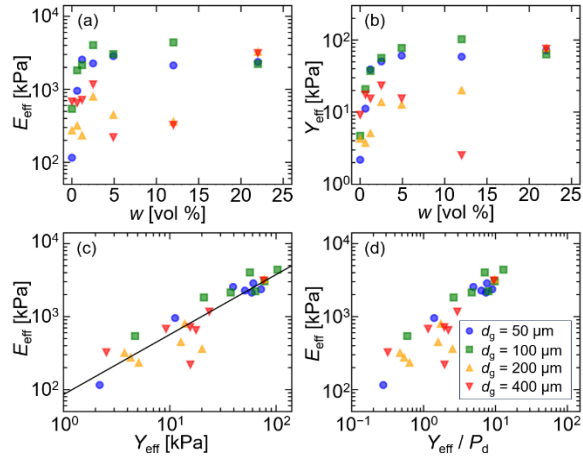

**FIG. 6.** Effects of the water content $w$ on (a) the effective elasticity $E_{eff}$ and (b) the effective strength $Y_{eff}$. Correlations between $E_{eff}$ and (c) $Y_{eff}$, and (d) $Y_{eff} / P_d$ ($P_d = 0.5\, \rho_w U_0^2$).

## IV. DISCUSSION

In the previous section, we described various observed phenomena as well as the mechanical characteristics of the substrate that stems from the different substrate conditions. In this section, we discuss the phenomena in more detail to understand the mechanisms and to provide some quantitative predictions.

### A. Particle ejection and crater generation

Generation of the ejecta particles can be considered as a result of the substrate fracture. In other words, particles can remain inside the substrate if the effective strength of the substrate is sufficiently large, as we have similar experience of taking a walk on wet sands in a beach. In the present study, the threshold of the fracture is related to the dynamic pressure of the impact $P_d$. Figure 7a shows a relationship between the water content $w$ and effective strength normalized by $P_d$, in which the particle diameter is indicated by different set of symbol and color, while filled symbols indicate the case that the particle ejection was observed. We found that the particle ejection occurs when $Y_{eff} / P_d < 3$. Moreover, the threshold of $Y_{eff} / P_d = 3$ is also related to the maximum depth of the crater $z_c$ as shown in Fig. 7b. Figure 7b indicates that the degree of penetration depends on the balance between $Y_{eff}$ and $P_d$ when $Y_{eff} / P_d < 3$, whereas it is almost independent when $Y_{eff} / P_d > 3$. These results imply that the reaction of the substrate changes across the threshold: the fracture of the substrate occurs and particles are ejected as a result of the kinetic energy transfer in case of $Y_{eff} / P_d < 3$, whereas the particles are pressed downward but not ejected from the substrate by either forming closer packing and/or tightly holding each other through the capillary bridge for $Y_{eff} / P_d > 3$.

      Similar trend is also seen in the crater radius normalized by the initial drop radius $R_c / R_0$ ($R_c / R_0$ is

large for small $Y_{eff} / P_d$ and it takes almost constant for $Y_{eff} / P_d > 3$, see Fig. 8a). We also find that the results agree with a scaling of $R_c / R_0 \sim (P_d / Y)^{1/5}$ proposed by Zhang et al.[24], which was confirmed for $w < 0.5$ vol % with $d_g = 90$ μm ($Y$ denotes the shear yield stress of the substrate). Because their experimental condition ranges from $0.8 < P_d / Y < 25$, we consider that the fracture occurred in their study. Contrary, the fracture occurred only a half cases ($P_d / Y_{eff} > 1 / 3$) in the present study. However, Fig. 8b shows that $R_c / R_0$ still seems to obey the scaling for $P_d / Y_{eff} < 1 / 3$. Although Fig. 8a implies that $R_c / R_0$ would take a constant for lower $P_d / Y_{eff}$, further work is necessitated to fully understand the $P_d$ dependence of the crater radius.

The above discussion suggests that the boundaries of the ejecta generation and the crater Type (I / II) are given by the balance between the dynamic pressure and effective strength of the substrate. However, the boundary for crater Type (II / III) cannot be explained by $Y_{eff} / P_d$. The observation results imply that the Type III crater is formed when the drop penetrates as deep as its apex reaches to the substrate surface.

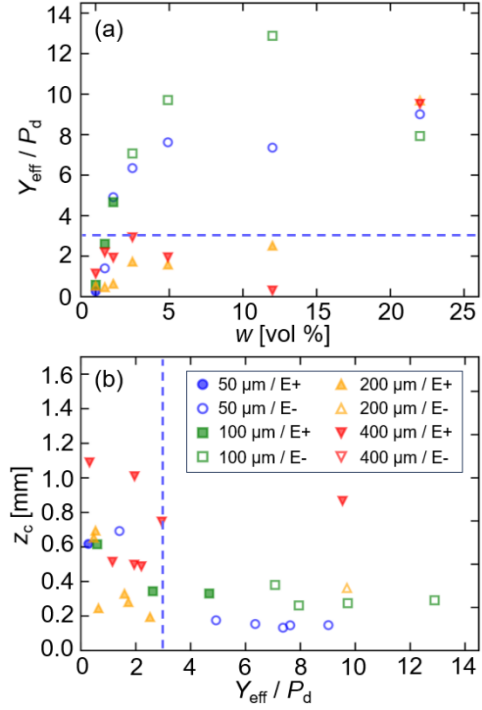

**FIG. 7. Relationships between (a) the water content $w$ and normalized effective strength $Y_{eff} / P_d$, and (b) $Y_{eff} / P_d$ and $z_c$. Filled symbols indicate the cases that the ejecta particles were observed ("E+" and "E-" in legend indicate the cases with and without ejecta, respectively). Blue dashed lines indicate $Y_{eff} / P_d = 3$.**

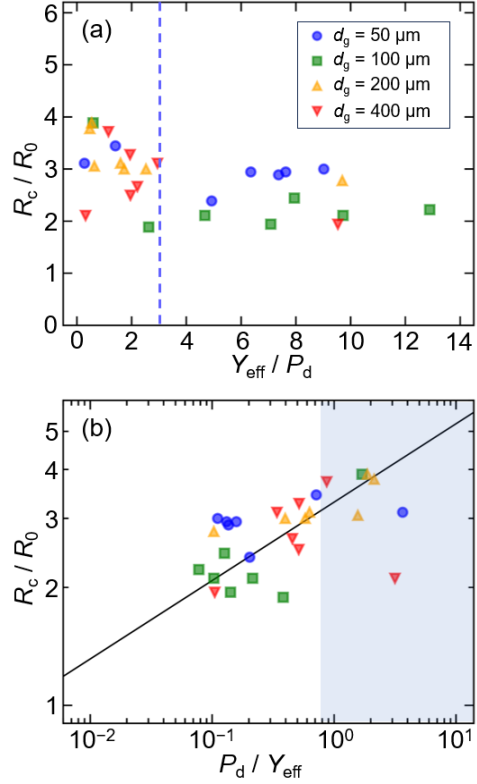

**FIG. 8.** (a) A relationship between the normalized effective strength $Y_{eff}/P_d$ and crater radius normalized by the initial drop radius $R_c/R_0$. Blue dashed line indicates $Y_{eff}/P_d = 3$. (b) A log-log plot of $P_d/Y_{eff}$ and $R_c/R_0$. Black line and blue-grayed area indicate $R_c/R_0 = 3.3\,(P_d/Y_{eff})^{1/5}$ and the range where $R_c/R_0 \sim (P_d/Y_{eff})^{1/5}$ scaling was confirmed by Zhang et al.[24].

### B. Splashing

We discussed in the previous subsection that the boundary of crater Type (I / II) is determined by $Y_{eff}/P_d$. As the boundary is common for that of splashing between Phase 2 and Phase 3, we can understand that the energy dissipation by the substrate fracture is substantially reduced in Phase 2 (and Phase 1). On the other hand, in Phase 3, penetration into the substrate occurs but still weak receding splash is observed while no splashing phase (Phase 4) also exists. In the following part, we discuss the effects of roughness, elasticity, and surface deformation on the suppression of the splashing.

*Effects of roughness*

It is known that the prompt splash is promoted on rough substrates[13,14,16,38]. However, the substrates in this study seemed to behave differently (although surface becomes rough as $d_g$ increases as shown in Table 1, the splash tends to be suppressed for large $d_g$ as shown in Fig. 2). It is because we define the splash as the release of relatively large droplets (> 100 μm), whereas the previous studies focused on smaller droplets (≤ 100 μm). As mentioned earlier, we also observed these small droplets ejected in very early stage of the event even with $d_g$ = 400 μm, which provides significant roughness. It implies that the roughness effect is still consistent in this study, but the boundary of the onset of the prompt splash (in our definition) is different from the previous studies. Subsequently, we discuss this mechanism.

*Effects of elasticity*

The suppression of the splashing could be owing to the energy dissipation during the spreading regime. A possible cause of the dissipation is that due to the substrate elasticity. Howland et al.[7] reported that the elasticity of the substrate $E$ affects the splashing on it, and the threshold of $We$ for the splash is raised when $E$ is decreased. Similar result is also reported by Basso et al.[37], although the threshold values are different from those reported by Howland et al. [7]. While these studies suggest that the elasticity affects the splash, we rule out the elasticity effect because of the following reasons: (i) high $We$ (= 395) of this study, at which the splash is predicted when $E \geq 10^3$ Pa in both study, (ii) $E_{eff}$ and $Y_{eff}$ have a positive correlation and $E_{eff} \geq 10^6$ Pa when $Y_{eff}/P_d > 3$ (Fig.

6d), namely, fracture occurs in the "elastic" region, and (iii) literature[5,6,37,39] report that the maximum spreading diameter is not affected by the substrate elasticity in our experimental range ($E > 10^2$ kPa) whereas the spreading diameter is obviously affected by the substrate condition in our case (Fig. 1).

*Effects of surface deformation*

The fracture of substrate occurs when the strength of the substrate is below the threshold of $Y_{eff} / P_d \sim 3$. Our result suggests that $Y_{eff} / P_d$ can also be used for determining the onset of the prompt splash. Figure 9 shows that the prompt splash occurs (Phase 1 and 2) above the threshold of $Y_{eff} / P_d \sim 3$. However, it also suggests that the boundary between the weak splashing (Phase 3) and no splashing (Phase 4) cannot be explained by the fracture of the substrate. We will discuss on this boundary in the next subsection.

The boundary between Phase (1 / 2) may be given by the deformed surface profiles: the deformation of the substrate changes the spreading direction from horizontal to obliquely upward (typically $10^1$ degrees[22]). The change in the spreading direction leads the spreading liquid front to take off the substrate at $r > R_c$ and to deform the interface by the Rayleigh–Plateau instability without the energy loss due to the liquid–solid contact. We qualitatively observed that the take-off angle of lamella in Phase 1 and 2 decreased with $w$ (Fig. 1).

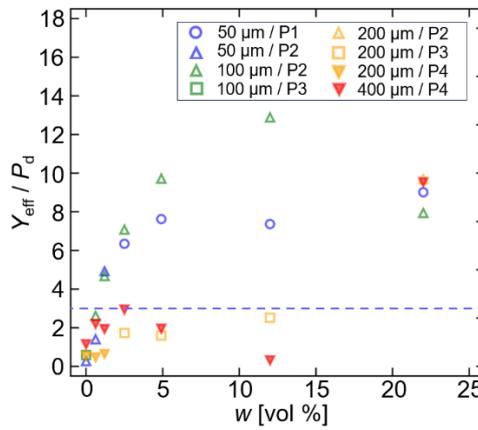

FIG. 9. A relationship between the water content $w$ and normalized effective strength $Y_{eff} / P_d$. Color indicates the grain diameter (blue: $d_g = 50$ μm, green: $d_g = 100$ μm, orange: $d_g = 200$ μm, red: $d_g = 400$ μm) and symbol indicates the splashing phase (open circle: Phase 1, open upper triangle: Phase 2, open square: Phase 3, filled lower triangle: Phase 4). Blue dashed line indicates $Y_{eff} / P_d = 3$.

### C. Event timescales

From the above discussion, we understand that Phase 3, where release of few large droplets covered by grains were observed, is a transition region between the splashing and penetration. In this region, $Y_{eff} / P_d$ is slightly less than the threshold and therefore the penetration occurs. However, although the viscous dissipation in the granular layer becomes nonnegligible[40], the impacting drop still expands in horizontal direction while trapping the surrounding grains on its interface. Finally, the release of the droplets occurs during the receding process of the interface.

To quantitatively evaluate the above scenario, we derive several timescales and compare them. First, the timescale of impact, which represents the early stage of the event, is derived as $\tau_i \sim R_0/U_0$. Second, the timescale of contact, which is derived from a relationship between the impact energy and surface energy after the drop deformation, is derived as $\tau_c \sim (\rho_w R_0^3/\gamma)^{1/2}$. In this study, $\tau_i \sim 0.5$ ms and $\tau_c \sim 8.9$ ms are obtained. In addition to these timescales, we introduce the third timescale $\tau_v$. This timescale represents the time at which the direction of the drop motion changes from vertical to horizontal, and explains the horizontal expansion after the penetration. We derive the timescale from the stopping force of the vertical motion. The candidates are capillarity and viscous force.

Here, we consider the impact of a drop onto a hydrophilic substrate which has dense cylindrical pores (pore radius of $r_p$). To stop the motion, the capillary pressure near the entrance of the pores ($\sim \gamma/r_p$) has to exceed the dynamic pressure ($\sim \rho_w U_0^2$), but it is one order smaller than the counterpart in this study when we put $r_p = d_g/2$. Therefore, we consider that penetration occurs even when $d_g = 50$ μm.

Now, our concern is that how deep (long) the liquid penetrates into the substrate. In the early stage of the penetration, the viscosity effect is negligible and inertia drives the liquid motion[41]. This stage ends as the viscous boundary layer grows, and this time is given by $\tau_v = \rho_w r_p^2 / 4\eta$, at which viscous stress inside the pore is nonnegligible. As the viscous stress increases inside the pore, velocity of the liquid front decreases and the drop expands in horizontal direction to satisfy the mass conservation. The resistance pressure due to the viscous effect is estimated as $P_v \sim \eta U_0 z(t) / r_p^2$, where $z(t)$ is the penetration depth inside the pore and scaled as $z(t) \sim U_0 t$. Therefore, $P_v$ is rewritten as:

$$P_v \sim \eta U_0^2 t / r_p^2 \qquad (3)$$

Substitution of $\tau_v$ into Eq. (3) yields $P_v \sim \rho_w U_0^2$, which is the same order to the dynamic pressure and suggests that $P_v$ is sufficiently large to stop the vertical motion.

As the newly introduced timescale is proportional to $r_p^2$ ($\sim d_g^2$), $\tau_v$ varies almost two orders between $d_g$ = 50 μm (~0.2 ms) and 400 μm (~10 ms), and $\tau_v$ becomes longer than $\tau_c$ for $d_g$ = 400 μm. It suggests that the liquid penetrates into vertical direction and does not expand in horizontal direction for $d_g$ = 400 μm, while the change in direction occurs at some point before the drop completely penetrate into the substrate for $d_g \leq 200$ μm.

The penetration affects the drop behavior even for $d_g \leq 200$ μm. For $d_g$ = 50 μm, $\tau_v < \tau_i$ and therefore the effect of the penetration can be negligible. However, for $d_g \geq 100$ μm, $\tau_v$ becomes longer than $\tau_i$ ($\tau_v \sim 0.6$ ms for $d_g$ = 100 μm, $\tau_v \sim 2.5$ ms for $d_g$ = 200 μm). In this case, a part of liquid penetrates into the substrate and the horizontal expansion attenuates. This attenuation leads a delay or a disappearance of the splashing. This trend can be seen in Fig. 10, which shows the onset time of the splashing $t_{sp}$ normalized by the timescale of contact $\tau_c$. It also shows that a threshold between Phase (2 / 3) is given by $t_{sp} / \tau_c \sim 0.5$. It is natural because the splashing mode is different and $\tau_c$ characterizes the timescale of drop expansion: the prompt splash releases droplets during the expanding stage while the receding splash releases them during the retracting stage.

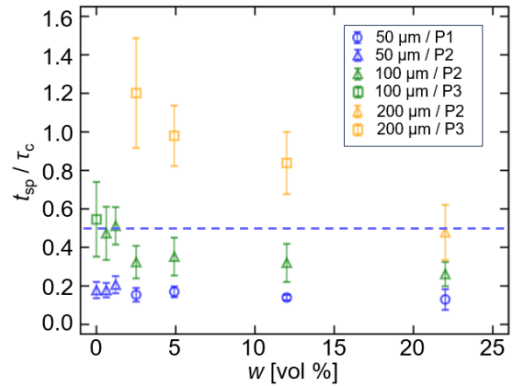

**FIG. 10. Measured time of the onset of splashing $t_{sp}$ normalized by the contact timescale $\tau_c$. Color indicates the grain diameter (blue: $d_g$ = 50 μm, green: $d_g$ = 100 μm, orange: $d_g$ = 200 μm) and symbol indicates the splashing phase (open circle: Phase 1, open upper triangle: Phase 2, open square: Phase 3). Blue dashed line indicates $t_{sp} / \tau_c = 0.5$.**

## V. CONCLUSIONS

We investigated the drop impact onto dry/wet granular substrates to understand the mechanisms of liquid splashing and cratering. Although we fixed the impactor condition (impact velocity: 4.0 m s$^{-1}$, drop radius: 1.8 mm), observations with different grain diameter (50–400 μm) and water content (0–22 vol %) exhibited rich variations in splashing mode, crater morphology, and particle ejection. We sorted these variations into four splashing phases, three crater-shape types, and existence of the particle ejection. Comparison of the phase diagrams for the splashing phase and crater-shape type revealed that the splashing and the cratering (as well as the particle ejection) are related each other.

To quantitatively understand the physics behind the variations, we measured mechanical characteristics (roughness, effective elasticity, and effective strength) of the substrates and characteristic lengths of the craters. Consequently, we concluded that the effective strength is a key parameter for both the splashing mode and crater morphology. In cases that the effective strength normalized by the dynamic pressure of the impacting drop, $Y_{eff} / P_d$, is smaller than a threshold (approximately three in this study), fracture of the substrate occurs and the particles are ejected. On the other hand, only little substrate deformation occurs in the cases that $Y_{eff} / P_d$ exceeds the threshold.

The difference in the response of the substrate affects both the forthcoming liquid spreading and the final crater shape. The liquid penetration inhibits the drop spreading and the following splashing (splashing Phase 4). In such cases, the resulting crater shape is a bowl-like (crater Type III). The drop starts to splash as $Y_{eff} / P_d$ reaches to the level of the threshold. At this point, the splashing is modest because the fracture still occurs (weak receding splash, splashing Phase 3). In this case, small droplets generated by the splashing are covered by the particles because the liquid penetration also occurs simultaneously and the drop shoves the surrounded particles while spreading. The shoved particles attach onto the drop interface, and some of them are released with the splashing droplets, while the rest of them are attracted to the crater center by the surface tension of liquid and form a dome-like shape (crater Type II). The splashing is intense (prompt splashing) when $Y_{eff} / P_d$ exceeds the threshold (splashing Phase 2). In such cases, the resulting crater shape is almost flat and the substrate deformation in vertical direction is only few particle diameters (crater Type I). We qualitatively observed that the angle of splash depends on the crater depth in this regime, and the angle is almost horizontal when the substrate is wet and composed of 50-μm grains (splashing Phase 1).

Finally, we discussed on three timescales of the event to qualitatively understand the thresholds for Phase (1 / 2) and Phase (3 / 4) of splashing. Comparison of the timescales of impact, contact, and penetration suggested that the shortest timescale dominates the event. In particular, we found that the timescale of penetration, which is a quadratic function of the grain diameter ($\tau_v \sim \rho_w d_g^2 / 4\eta$), significantly affects the event because it also indicates the timescale of the change in the direction of motion from vertical (penetration) to horizontal (spreading). Although the above discussion on the timescale neglects the effects of the water content, our observation results imply that the impact event becomes more complicated when the substrate is wet, even when it is slightly moist. It could be due to the existence of liquid inside the substrate that could shorten the penetration timescale.

In this study, we showed that both liquid splashing and crater morphology have wide varieties even when the impactor condition is the same. Our findings also suggest that the splashing and cratering show different responses when the impactor is a solid-like or an aggregation of grains such as a meteorite. More detailed penetration into the phenomena would extend our understanding in a wide range, from our daily-life problems (such as soil erosion or egg drop on flour) to planetary problems.

## AUTHOR CONTRIBUTIONS

**Wei Zhang:** Formal Analysis (equal), Investigation (lead), Methodology (supportive), Software (equal), Validation (lead), Visualization (lead), Writing-Review & Editing (equal)

**Hiroaki Katsuragi:** Conceptualization (equal), Data Curation (equal), Funding Acquisition (equal), Investigation (supportive), Methodology (supportive), Project Administration (equal), Resources (lead), Supervision (equal), Writing–Original Draft Preparation (supportive), Writing–Review & Editing (equal)

**Ken Yamamoto:** Conceptualization (equal), Data Curation (equal), Formal Analysis (equal), Funding Acquisition (equal), Investigation (supportive), Methodology (lead), Project Administration (equal), Software (equal), Supervision (equal), Writing–Original Draft Preparation (lead), Writing–Review & Editing (equal)

## CONFLICTS OF INTEREST

There are no conflicts to declare.

## ACKNOWLEDGMENTS

This study was financially supported by the Japan Society for the Promotion of Science (JSPS) KAKENHI Grant No. 23K17729 and 23H04134.